\documentclass[11pt]{article}

\usepackage[paper=a4paper,dvips,top=2.0cm,left=2.2cm,right=2.2cm,bottom=2.9cm,footskip=1.4cm,voffset=1.2cm]{geometry}
\usepackage{amsfonts,amsmath,amsthm,amssymb}
\numberwithin{equation}{section}  %%% Changes the equation numbering according to the section number
\usepackage[svgnames]{xcolor}
\usepackage{fix-cm}
\usepackage{sectsty}
\usepackage{fancyhdr}
\usepackage{lastpage}
\usepackage{graphicx}
\usepackage{url}
\usepackage{enumerate}
\usepackage{paralist}
\usepackage{multicol}
\usepackage{color}  %%% Needs to make colour of some content of your text
\usepackage{float}  %%% Fixes the position of the figure
\usepackage{epsfig}
\usepackage{hyperref}   %%% Needs to make hyper reference of any references or citations
\hypersetup{colorlinks=true,linkcolor=beamer@PRD, citecolor=beamer@PRD}
\usepackage{authblk}  %%% Helps to arrange author and institution details.
\usepackage{cite}  %%% Colour the citation brackets

\newcommand\myref[1]{\textcolor{beamer@PRD}{(}\ref{#1}\textcolor{beamer@PRD}{)}}

\definecolor{beamer@blue}{RGB}{0,0,255}
\definecolor{beamer@mediumblue}{RGB}{0,0,190}
\definecolor{beamer@midnightblue}{RGB}{25,25,112}
\definecolor{beamer@navy}{RGB}{0,0,128}
\definecolor{beamer@darkblue}{RGB}{0,0,139}
\definecolor{beamer@purple}{RGB}{128,0,128}
\definecolor{beamer@levander}{RGB}{100.,149.,237.}
\definecolor{beamer@PRD}{RGB}{46,48,146}
\definecolor{beamer@green}{RGB}{0,128,0}
\definecolor{beamer@darkgreen}{RGB}{0,100,0}
\definecolor{beamer@olive}{RGB}{128,128,0}
\definecolor{beamer@darkolivegreen}{RGB}{85,107,47}
\definecolor{beamer@gray}{RGB}{190,190,190}
\definecolor{beamer@ivry}{RGB}{220,220,220}%{238,232,205}
\definecolor{beamer@new}{RGB}{40,120,50}
\definecolor{shadecolor}{RGB}{220,220,220}
\definecolor{beamer@darkslategray}{RGB}{47,79,79}
\definecolor{beamer@chocolate}{RGB}{210,105,30}
\definecolor{beamer@brown}{RGB}{165,42,42}
\definecolor{beamer@orangered}{RGB}{255,69,0}
\definecolor{beamer@maroon}{RGB}{128,0,0}
\definecolor{beamer@white}{RGB}{234,242,243}
\definecolor{beamer@silver}{RGB}{0.5,0.45,0.37}
\begin{document}

%%%%%%%%%%%%%%%%%%%%%%%%%%%%%%%%%%%%%%%%%%%%%%%%%%%%%%%%%%%%%%%%%%%%%%%%%%%%%%
%  Title
%%%%%%%%%%%%%%%%%%%%%%%%%%%%%%%%%%%%%%%%%%%%%%%%%%%%%%%%%%%%%%%%%%%%%%%%%%%%%%
\title{\textbf{Modification of Schr\"odinger-Newton equation due to braneworld models with minimal length}}
\author{\textbf{Anha Bhat$^\bullet$, Sanjib Dey$^\ast$, Mir Faizal$^\circ$, Chenguang Hou$^\dagger$ and Qin Zhao$^\dagger$} \\ \footnotesize{$^\bullet$Department of Metallurgical and Materials Engineering, National Institute of Technology, Srinagar 190006, India \\ $^\ast$Institut des Hautes \'Etudes Scientifiques, Bures-sur-Yvette 91440, France $\&$\\ Institut Henri Poincar\'e, Paris 75005, France \\ $^\circ$Irving K. Barber School of Arts and Sciences, University of British Columbia-Okanagan, \\ 3333 University Way, Kelowna, BC V1V 1V7, Canada $\&$ \\ Department of Physics and Astronomy, University of Lethbridge, Lethbridge, AB T1K 3M4, Canada \\ $^\dagger$Department of Physics, National University of Singapore, 2 Science Drive 3, Singapore \\ E-mail: anhajan1@gmail.com, sdey@ihes.fr, mirfaizalmir@googlemail.com, \\ a0086303@u.nus.edu, a0086307@u.nus.edu}}
\date{}
\maketitle
%%%%%%%%%%%%%%%%%%%%%%%%%%%%%%%%%%%%%%%%%%%%%%%%%%%%%%%%%%%%%%%%%%%%%%%%%%%%%%
%  Abstract
%%%%%%%%%%%%%%%%%%%%%%%%%%%%%%%%%%%%%%%%%%%%%%%%%%%%%%%%%%%%%%%%%%%%%%%%%%%%%%
\thispagestyle{fancy}
\begin{abstract}
We study the correction of the energy spectrum of a gravitational quantum well due to the combined effect of the braneworld model with infinite extra dimensions and generalized uncertainty principle. The correction terms arise from a natural deformation of a semiclassical theory of quantum gravity governed by the Schr\"odinger-Newton equation based on a minimal length framework. The two fold correction in the energy yields new values of the spectrum, which are closer to the values obtained in the GRANIT experiment. This raises the possibility that the combined theory of the semiclassical quantum gravity and the generalized uncertainty principle may provide an intermediate theory between the semiclassical and the full theory of quantum gravity. We also prepare a schematic experimental set-up which may guide to the understanding of the phenomena in the laboratory. 
\end{abstract}	 
%%%%%%%%%%%%%%%%%%%%%%%%%%%%%%%%%%%%%%%%%%%%%%%%%%%%%%%%%%%%%%%%%%%%%%%%%%%%%%
%  Introduction
%%%%%%%%%%%%%%%%%%%%%%%%%%%%%%%%%%%%%%%%%%%%%%%%%%%%%%%%%%%%%%%%%%%%%%%%%%%%%%
\section{Introduction} \label{sec1}
\addtolength{\footskip}{-0.2cm} %% Adds extra space in the first page of the article
The fact that the minimal observable length can be useful to impose an effective cut-off in the ultraviolet domain in order to make the theory of quantum fields renormalizable was suggested very early by Heisenberg. It was Snyder, who formalized the idea for the first time in the form of an article, and showed that the noncommutative structure of space-time characterizes the minimal measurable length in a very natural way \cite{Snyder}. Since then, the notion of noncommutativity has evolved from time to time and revealed its usefulness in different contexts of modern physics \cite{Seiberg_Witten}. Some natural and desirable possibilities arise when the canonical space-time commutation relation is deformed by allowing general dependence of position and momentum \cite{Kempf_Mangano_Mann,Das_Vagenas, Gomes,Bagchi_Fring,Quesne_Tkachuk}. In such scenarios, the Heisenberg uncertainty relation necessarily modifies to a generalized version to the so-called \textit{generalized uncertainty principle} (GUP). Over last two decades, it is known that within this framework, in particular, where the space-time commutation relation involves higher powers of momenta, explicitly lead to the existence of nonzero minimal uncertainty in position coordinate, which is familiar as the \textit{minimal length} in the literature. An intimate connection between the gravitation and the existence of the fundamental length scale was proposed in \cite{Mead}. The string theory, which is the most popular approach to quantum gravity, also supports the presence of such minimal length \cite{Veneziano,Gross_Mende,Amati,Yoneya,Konishi}, since the strings are the smallest probes that exist in perturbative string theory, and so it is not possible to probe space-time below the string length scale. In loop quantum gravity, the existence of a minimum length has a very interesting consequence, as it turns the big bang into a big bounce \cite{Rovelli,Dzierzak}. Furthermore, the arguments from black hole physics suggest that any theory of quantum gravity must be equipped with a minimum length scale \cite{Maggiore,Park}, due to the fact that the energy required to probe any region of space below the Plank length is greater than the energy required to create a mini black hole in that region of space. The minimal length exists in other subjects too; such as, path integral quantum gravity \cite{Padmanabhan,Greensite}, special relativity \cite{Amelino}, doubly special relativity \cite{Magueijo,Girelli,Cortes}, etc. In short, the existence of minimal measurable length, by now, has become a universal feature in almost all approaches of quantum gravity. Moreover, some Gedankenexperiments and
thought experiments \cite{Mead,Scardigli,Ng} in the spirit of black hole physics have also supported the idea. For further informations on the subject one may follow some review articles devoted to the subject, for instance, \cite{Garay,Ng_Review,Hossenfelder_Review}.

In spite of having several serious proposals for quantizing the general relativity, unfortunately we do not have a fully consistent quantum theory of gravity yet. This has motivated the study of \textit{semi-classical quantum gravity} (SCQG), where the gravitational filed is treated as a background classical field, and the matter fields are treated quantum mechanically. Under such approximation, if $|\psi\rangle$ is the wave function of the matter field, the Einstein tensor $G_{\mu\nu}$ can be obtained in terms of the quantum mechanical energy-momentum tensor for matter fields $T^{\mu\nu}$ as $G_{\mu\nu} =8\pi G\langle \psi|T^{\mu\nu}|\psi\rangle/c^4$. However, Newtonian gravity has been observed to be the correct approximation to general relativity till the smallest length scale (0.4mm) to which general relativity has been tested \cite{Yang}. Thus, at small distances, it is expected that the semi-classical approximation can be described by a Schr\"odinger-Newton equation \cite{Ruffini,Penrose,Tod,Van,Yang_Miao}, which is the nonrelativistic limit of the Dirac equation and the Klein-Gordon equation with a classical Newtonian potential \cite{Giulini}. It has been proposed that the Schr\"odinger-Newton equation can be utilized to explore various interesting properties of gravitational systems at the given length scale \cite{Giulini_Brobardt,Manfredi,Bera}.

Some interesting consequences may follow from the combination of the above two frameworks, namely, the GUP and SCQG, and it is allowed since both of the effects occur at small scales. Technically, this can be achieved by imposing the minimal length structure into the semi-classical scheme of gravity by means of deforming the Schr\"odinger-Newton equation in accordance with the laws of GUP. The most interesting fact is that the new theory resulting from the combination of GUP and SCQG may provide an intermediate theory between a full quantum theory of gravity and the SCQG and, this is precisely the issue that we address in the present manuscript. Recently, it has been suggested that both the Schr\"odinger-Newton equation \cite{Grossardt,Gan} and the deformation of quantum mechanical structure by the GUP \cite{Pikovski} can be tested experimentally by using the opto-mechanical systems. Therefore, it becomes important to understand the effects of the combined theory, especially when there is a strong viability that the theory may be tested by using a similar type of experimental set-up in near future.
%%%%%%%%%%%%%%%%%%%%%%%%%%%%%%%%%%%%%%%%%%%%%%%%%%%%%%%%%%%%%%%%%%%%%%%%%%%%%%
% Section 2
%%%%%%%%%%%%%%%%%%%%%%%%%%%%%%%%%%%%%%%%%%%%%%%%%%%%%%%%%%%%%%%%%%%%%%%%%%%%%%
\section{GUP and minimal length}\label{sec2}
\lhead{Modified Schr\"odinger-Newton equation}
\chead{}
\rhead{}
\addtolength{\voffset}{-1.2cm} %% Adds extra space in the header of the first page of the article 
\addtolength{\footskip}{0.2cm} %% Adds extra space in the first page of the article 
Let us commence by introducing a particular version of modified commutation relation between the position and momentum operators $X,P$ \cite{Kempf_Mangano_Mann,Das_Vagenas}
\begin{equation}\label{GUP}
[X_i, P_j] = i\hbar(\delta_{ij} + \beta\delta_{ij}P^2 + 2\beta P_i P_j), \qquad [X_i,X_j]=[P_i,P_j]=0,
\end{equation}
with $\beta$ having the dimension of inverse squared momentum and, $P^2=\sum_{j=1}^{3}P_j^2$. As obviously, in the limit $\beta\rightarrow 0$, the deformed relations \myref{GUP} reduce to the standard canonical commutation relations $[x_i,p_j]=i\hbar\delta_{ij}$. However, there exist many other similar type of deformations in the literature, which have been used to investigated many interesting phenomena in different contexts; see, for instance \cite{Bagchi_Fring,Quesne_Tkachuk,Pedram_Nozari_Taheri, Dey_Fring_Gouba,Hossenfelder_Review,Dey_Hussin}. Nevertheless, the generalized uncertainty relation or the Robertson-Schr\"odinger uncertainty relation corresponding to the deformed algebra \myref{GUP} turns out to be
\begin{eqnarray}
&& \Delta X_i\Delta P_j \geq \frac{1}{2}\Big\vert\langle[X_i,P_j]\rangle\Big\vert \\
&& \Delta X_i\Delta P_i \geq \frac{\hbar}{2}\left[1+\beta\left\{(\Delta P)^2+\langle P\rangle^2\right\}+2\beta\left\{(\Delta P_i)^2+\langle P_i\rangle^2\right\}\right],
\end{eqnarray}
from which one can compute the exact expression of the minimal observable length by using the standard minimization technique. For further information on the minimization procedure, one may refer, for instance \cite{Kempf_Mangano_Mann,Bagchi_Fring,Dey_Fring_Gouba}. A possible representation of the algebra \myref{GUP} in terms of the canonical position and momentum operators $x,p$ is given by \cite{Kempf_Mangano_Mann,Das_Vagenas}
\begin{equation}\label{Rep}
X_i = x_i, \qquad P_i = p_i(1+\beta p^2), 
\end{equation}
where $p_j = -i\hbar\frac{\partial}{\partial {x}_j}$. 
Certainly, the representation \myref{Rep} is not unique \cite{Kempf_Mangano_Mann},
as for instance; see \cite{Dey_Fring_Khantoul}, where the authors explore four
possible representations of the algebra \myref{GUP} with some of them being Hermitian. However, it is easy to show that \myref{Rep} satisfies \myref{GUP} up to the first order of $\beta$ (hence, we neglect higher order terms of $\beta$). Physically, the notations are understood in the sense that $p_i$ represents the momenta at low energy, while $P_i$ correspond to those at high energy.
%%%%%%%%%%%%%%%%%%%%%%%%%%%%%%%%%%%%%%%%%%%%%%%%%%%%%%%%%%%%%%%%%%%%%%%%%%%%%%
% Section 3
%%%%%%%%%%%%%%%%%%%%%%%%%%%%%%%%%%%%%%%%%%%%%%%%%%%%%%%%%%%%%%%%%%%%%%%%%%%%%%
\section{The gravitational quantum well}\label{sec3}
The gravitational quantum well (GQW) is characterized by the motion of a nonrelativistic object of mass $m$ in a gravitational field $\overrightarrow{g}=-g\overrightarrow{e}_z$ with a restriction imposed by a  mirror placed at the origin $z=0$, such that the potential turns out to be
\begin{align}\label{GravPot}
\begin{aligned}
V_0(z) =\left\lbrace\begin{array}{l}
 +\infty,  ~~~z\leq 0, \\
 mgz, ~~~z>0,
\end{array} \right.
\end{aligned}
\end{align}
with $g$ being the gravitational acceleration. The experimental set-up of the corresponding potential \myref{GravPot} has already been studied \cite{Nesvizhevsky}. Theoretically, the problem resemble a quantum mechanical particle moving in a potential well given by \myref{GravPot} subjected to a boundary condition $\psi^0(0)=0$ at $z=0$. If we consider $\psi^0(\overrightarrow{x})=\psi^0(z)\psi^0(y)$, then the solutions of the corresponding time-independent Schr\"odinger equation along the $z$ direction   
\begin{equation}\label{Ham0}
H_0\psi^0(z) = E^0\psi^0(z), \qquad H_0 = \frac{p^2}{2m} +mgz,
\end{equation}
are well-known and, are given by \cite{Flugge} (Problem $40$)
\begin{eqnarray}
E_{n}^0 &=& -\frac{mg\alpha_n}{\theta}, \qquad \theta = \left(\frac{2m^2g}{\hbar^2}\right)^{1/3},\\
\psi_n^0(z) &=& N_n \text{Ai}[\theta z+\alpha_n],
\end{eqnarray}
where $\alpha_n$ is the $n$\textsuperscript{th} zero of the regular Airy function $\text{Ai}(z)$ \cite{Abramowitz_Stegun}, and $N_n=\theta^{1/2}/|\text{Ai}'(\alpha_n)|$ is the normalization factor. Along the $y$ axis, the particle is free and the corresponding wave function takes the form
\begin{equation}
\psi^0(y)=\int_{-\infty}^{\infty}g(k)e^{iky}dk,
\end{equation}
where $g(k)$ determines the shape of the wave packet in momentum space. In analogy to a classical object, a collision of a quantum particle with the impenetrable mirror along the $z$ direction will make it bounced at a critical height $h_n$ 
\begin{equation}\label{hight}
h_n = \frac{E_n^0}{mg} = -\frac{\alpha_n}{\theta},
\end{equation}
which is, naturally, quantized. In the GRANIT experiment \cite{Nesvizhevsky1}, the measured critical heights for the first two states are 
 \begin{align}
 & h^{\text{exp}}_1 = 12.2\mu m \pm 1.8_{\text{syst}} \pm 0.7_{\text{stat}},\\
  & h^{\text{exp}}_2 = 21.6\mu m \pm 2.2_{\text{syst}} \pm 0.7_{\text{stat}}.
 \end{align}
Whereas, the theoretical values for $h_1$ and $h_2$ can be obtained from \myref{hight} as
\begin{align}
& h^{\text{th}}_1 = 13.7\mu m,\\
& h^{\text{th}}_2 = 24.0\mu m,
\end{align}
where $m=939~\text{Mev/c\textsuperscript{2}}$ and $g=9.81~\text{m/s\textsuperscript{2}}$. The variation of the heights $h_1$ and $h_2$ resulting from the theory and the experiment are, therefore, $\delta^{\text{th}}=h_2^{\text{th}}-h_1^{\text{th}}=10.3\mu m$ and $\delta^{\text{exp}}=h_2^{\text{exp}}-h_1^{\text{exp}}=9.4 \mu m \pm 5.4\mu m$, respectively. Thus, the deviation of $\delta$ between the theory and the experiment turns out to be $4.5\mu m$. The argument that we shall pursue here is that any correction in the Hamiltonian \myref{Ham0} will effectively reduce this deviation of $\delta$ between the theory and the experiment, and thus the theoretical result will become closer to the experiment. A similar argument has already been used in \cite{Bertolami_Rosa_etal,Brau,Banerjee,Buisseret}. However, we explore two types of correction here. First, we consider a braneworld model studied in \cite{Randall}, which was discovered in the course of solving the hierarchy problem of the standard model. Their theory is based on the assumption that, while all the standard model fields, gauge and matter, are confined in a (3+1) dimensional manifold, only the graviton can propagate freely in the extra dimensions which are considered to be infinite. In presence of such infinite extra dimensions, the Newtonian potential is modified to the following form
\begin{equation}\label{PotCor}
V(r)=-\frac{GMm}{r}\left(1+\frac{k_b}{r^b}\right), \quad r>>\Lambda =\sqrt[b]{|k_b|},
\end{equation}
where $\Lambda$ is the length scale at which the correction due to the infinite extra dimensions becomes dominant. If we consider the Newtonian potential $-GMm/r=V_0(r)$, we can write the above equation \myref{PotCor} as $V(r)=V_0(r)+V_b(r)$, so that $V_b(r)=-GMmk_b/r^{b+1}$ can be considered as a perturbation, which will eventually contribute to the correction over the theoretical values of $\delta$. Therefore, $\delta^{\text{th}}$ will be changed to $\delta^{\text{th}}+\Omega$, where
\begin{equation}\label{Omega}
\Omega=\frac{1}{mg}\left[\left\langle\psi_2^0(z)|V_b|\psi_2^0(z)\right\rangle -\left\langle\psi_1^0(z)|V_b|\psi_1^0(z)\right\rangle\right],
\end{equation}
and $|\Omega|\leq 4.5\mu m$. The second correction emerges from the contribution of the GUP deformed Hamiltonian. If we replace the momentum $p$ corresponding to the low energy in \myref{Ham0} by the momentum $P$ coming from the GUP deformation \myref{Rep}, we obtain
\begin{equation}\label{GUPCorr}
H=H_0+H_1, \qquad H_1=\frac{\beta}{m}p^4,
\end{equation}
where we neglect the higher order terms of $\beta$. Thus, if we denote the correction of $\delta^{\text{th}}$ coming from the GUP by $\Xi$, with
\begin{equation}
\Xi = \frac{1}{mg}\left[\left\langle\psi_2^0(z)|H_1|\psi_2^0(z)\right\rangle -\left\langle\psi_1^0(z)|H_1|\psi_1^0(z)\right\rangle\right],
\end{equation}
then
\begin{equation}
|\Omega|\leq|\Omega+\Xi|\leq 4.5\mu m.
\end{equation}
Let us now compute both of the corrections $\Omega$ and $\Xi$ in the following section.
%%%%%%%%%%%%%%%%%%%%%%%%%%%%%%%%%%%%%%%%%%%%%%%%%%%%%%%%%%%%%%%%%%%%%%%%%%%%%%
%  Section 4
%%%%%%%%%%%%%%%%%%%%%%%%%%%%%%%%%%%%%%%%%%%%%%%%%%%%%%%%%%%%%%%%%%%%%%%%%%%%%%
\section{Corrections}\label{sec4} 
The potential described in \myref{PotCor} corresponds to an interaction between two particles. However, we are dealing with a situation where a point particle bounces on a plane mirror placed at the surface of the Earth. Therefore, we are required to derive the effective potentials between the test particle and the Earth, as well as, between the particle and the mirror.
\subsection{Correction due to Earth-particle interaction} 
Let us first derive the potential coming from the Earth-particle interaction. By considering our planet to be a spherical body with mass density $\rho_E$, the total potential acting on the particle is 
\begin{align}
V_b^E(\overrightarrow{r}) = -mG\rho_E k_b\int_E\frac{d^3\overrightarrow{r}'}{|\overrightarrow{r}-\overrightarrow{r}'|^{b+1}}.
\end{align}
In spherical coordinates which becomes
\begin{align}
V_b^E(h) = -2\pi mG\rho_E k_b\int^R_0 r^2dr\int^{1}_{-1} \frac{du}{[r^2-2r(h+R)u+(h+R)^2]^{(b+1)/2}},
\end{align}
where $h$ is the altitude above Earth's surface and $R$ being the radius of the Earth. For, $b=0,1,2,3$, one obtains \cite{Buisseret}
\begin{align}\label{power-up}
\begin{aligned}
& V_0^E(h) = -\frac{GmM}{h+R}, \\
& V_1^E(h) = -\frac{3mgk_1}{4R}\left[2R - \frac{h(h+2R)}{h+R}\ln\left(\frac{h+2R}{h}\right)\right],\\
& V_2^E(h) = -\frac{3mgk_2}{2R}\left[ \ln\left(\frac{h+2R}{h}\right) - \frac{2R}{h+R}\right],\\
& V_3^E(h) = -\frac{3mgk_3}{4R}\left[\frac{2R}{h(h+2R)}- \frac{1}{h+R}\ln\left(\frac{h+2R}{h}\right)\right],
\end{aligned}
\end{align}
while for $b>3$, it becomes
\begin{equation}
V_b^E(h) = -\frac{3mgk_b}{2(b-1)(b-2)(b-3)R(h+R)}\left[\frac{(b-3)R-h}{h^{b-2}}+\frac{(b-1)R+h}{(h+2R)^{b-2}}\right].
\end{equation}
Notice that in the limit $b\rightarrow 0~(k_0=1)$, we recover the Newtonian potential in an exact form.
%%%%%%%%%%%%%%%%%%%%%%%%%%%%%%%%%%%%%
\subsection{Correction due to mirror-particle interaction}
Although the mass of the mirror is negligible with respect to that of the Earth, the effect originating from the mirror-particle interaction should be taken into account since the mirror is placed at a distance much closer to the particle. The mirror can be seen as a parallelepiped with density $\rho_M$, and located at $-\infty <(x,y)<\infty$ and $-L<z<0$. With respect to the size of the particle, the mirror can be considered as an infinite plane, so that the mirror behaves as a disc of infinite radius. Thus, the total interacting potential in this case becomes
\begin{align}\label{MirrorInt}
V_b^M(h) = -2\pi mG\rho_M k_b\int^0_{-L} dz\int^{R^*}_{0} \frac{rdr}{[r^2+(h-z)^2]^{(b+1)/2}},
\end{align}
where $R^*$ is taken to be very large, but finite, in order to avoid the divergent integrals. Nevertheless, the integrals \myref{MirrorInt} are computed as follows
\begin{align}\label{mirror}
\begin{aligned}
& V_1^M(h) = \frac{3mgk_1\rho}{2R}\left[L\ln(h+L)+h\ln\left(\frac{h+L}{h}\right)\right],\\
& V_2^M(h) = -\frac{3mgk_2\rho}{2R} \ln\left(\frac{h+L}{h}\right),\\
& V_{b>2}^M(h) = -\frac{3mgk_b\rho}{2R(b-1)(b-2)}\left[\frac{1}{h^{b-2}}+\frac{1}{(h+L)^{b-2}}\right].
\end{aligned}
\end{align}
%%%%%%%%%%%%%%%%%%%%%%%%%%%%%%%%%%%%%%
\subsection{The perturbative correction from GUP deformation}
Now, let us consider the  correction due to the perturbation term $H_1$ \myref{GUPCorr} arising from the GUP deformation. The correction to the energy $\Delta E_n$ at the lowest order in $\beta$ is given by
\begin{align}
\begin{aligned}
\Delta E_n^{\text{GUP}} & = \frac{\beta}{m}\left\langle\psi(\overrightarrow{x})|p^4|\psi(\overrightarrow{x})\right\rangle \\
& = \frac{\beta}{m}[4m^2\left\langle\psi_n^0(z)|p^4_z|\psi_n^0(z)\right\rangle + 2\left\langle\psi_n^0(z)|p^2_z|\psi_n^0(z)\right\rangle\left\langle\psi^0(y)|p^2_y|\psi^0(y)\right\rangle \\
& = \frac{\beta}{m}[4m^2\left\langle [E^0_n - V_0(z)]^2\right\rangle + 8m^2E_c\left\langle E^0_n-V_0(z)\right\rangle] \\
& = 4\beta m[E^0_n(E^0_n +2E_c) -2(E^0_n+E_c)\left\langle V_0(z)\right\rangle +\left\langle V_0^2(z)\right\rangle],
\end{aligned}
\end{align}
where $E_c = m\left\langle\psi^0(y)|v^2_y|\psi^0(y)\right\rangle /2$ is the kinetic energy of the particle along the horizontal direction. Note that a term proportional to $\left\langle\psi^0(y)|p^4_y|\psi^0(y)\right\rangle$ has been omitted since it only leads to a global shift of the energy spectrum. Therefore, by computing the following integrals
\begin{align}
\begin{aligned}
 \left\langle V_0(z)\right\rangle & =\langle\psi_n^0(z)|V_0(z)|\psi_n^0(z)\rangle= mgN_n^2\int^{+\infty}_0 z Ai^2(\theta z+\alpha_n) = \frac{2}{3}E_n^0, \\
 \left\langle V_0^2(z)\right\rangle & =\langle\psi_n^0(z)|V^2_0(z)|\psi_n^0(z)\rangle = (mg)^2N_n^2\int^{+\infty}_0 z^2 Ai^2(\theta z+\alpha_n) = \frac{8}{15}(E_n^0)^2,
\end{aligned}
\end{align}
we obtain the final expression of the correction to the energy as follows
\begin{align}
\Delta E_n^{\text{GUP}} = \frac{4}{5}\beta m(E^0_n)^2(1+\frac{10 E_c}{3 E^0_n}).
\end{align}
%%%%%%%%%%%%%%%%%%%%%%%%%%%%%%%%%%%%%%%%%%%%%%%%%%%%%%%%%%%%%%%%%%%%%%%%%%%%%%
%  Section 5
%%%%%%%%%%%%%%%%%%%%%%%%%%%%%%%%%%%%%%%%%%%%%%%%%%%%%%%%%%%%%%%%%%%%%%%%%%%%%%
\section{The modified GQW spectrum}\label{sec5}
Combining the corrections coming from the SCQG, we obtain 
\begin{equation}
V_b = V_b^E(z)+V^M_b(z),
\end{equation}
which when combined with the correction to the energy arising from the GUP deformation, we obtain the total energy shift
\begin{equation}
\Delta E_n =  \left\langle V_b\right\rangle +\Delta E^{\text{GUP}}_n.
\end{equation}
Therefore, we have 
\begin{equation}
\Omega_b + \Xi =\frac{1}{mg}\left[\left\langle\psi_2^0(z)|V_b|\psi_2^0(z)\right\rangle  + \Delta E_2^{\text{GUP}} - \left\langle\psi_1^0(z)|V_b|\psi_1^0(z)\right\rangle -\Delta E_1^{\text{GUP}}\right].
\end{equation}
Let us now compute the numerical values of the above expressions, with $R = 6378 \text{km}$, $m = 939 \text{MeV}/\text{c}^2, g = 9.81 \text{m}/\text{s}^2, L = 10\text{cm}, \rho= 1$ and $E_c \simeq m\left\langle  v_y\right\rangle^2/2 \simeq 0.221 \mu \text{eV}$
\begin{align}
& \Omega_1 +\Xi = k_1 \times 5.59203\times 10^{-11} + \beta\times1.62357\times 10^{-57}\text{m} < 4.5\mu \text{m}, \label{Eq54}\\
& \Omega_2 +\Xi = k_2 \times 2.38493\times 10^{-7} \text{m}^{-1} + \beta\times 1.62357\times 10^{-57}\text{m} < 4.5\mu \text{m}, \\
& \Omega_3 +\Xi = k_3 \times 0.0101239~\text{m}^{-2} + \beta\times1.62357\times 10^{-57}\text{m} < 4.5\mu \text{m}. \label{Eq56}
\end{align}
In order to obtain the bounds on the power-law parameters $\Lambda$ and $k_b$, let us consider $V_b=-mgk_b\tilde{V}_b$. Eq. \myref{PotCor} implies that $k_b = \Lambda^b$, so that $V_b=-mg\Lambda^b\tilde{V}_b$. Therefore, we can write
\begin{equation}
\Omega_b=-\Lambda^b\tilde{\Omega}_b=-\Lambda^b\left[\left\langle\psi_2^0(z)|\tilde{V}_b|\psi_2^0(z)\right\rangle - \left\langle\psi_1^0(z)|\tilde{V}_b|\psi_1^0(z)\right\rangle\right],
\end{equation}
which immediately leads to the constraint on the parameter $\Lambda$ as $|\Lambda|=\sqrt[b]{\Omega_b/\tilde{\Omega}_b}$. According to \myref{Omega}, $\Omega$ is bounded below $4.5\mu m$ and, so $\Lambda$ is bounded by the following relation
\begin{equation}
|\Lambda|=\sqrt[b]{4.5/\tilde{\Omega}_b}.
\end{equation}
\begin{table}
\centering 
\begin{tabular}{c|ccc}
$b$ & 1&2&3\\
\hline\\
$|\Lambda|$(m) $<$ & $8.13\times 10^{4}$ & $4.34$ & $0.076$
\end{tabular}
\caption{Constrains on the power-law parameters}
\label{constraint}
\end{table}
Using the results from \myref{Eq54}-\myref{Eq56}, we can calculate $\tilde{\Omega}_b$ corresponding to different values of $b$. If we turn off the GUP deformation ($\beta=0$), we obtain the Table \ref{constraint}. In a similar way it is also possible to obtain the bounds for the cases when $\beta\neq 0$. On the other hand, if we turn off the power-law modification, we see that the GUP deformation parameter has the constraint
\begin{equation}\label{Eq57}
\beta <2.77167\times 10^{51},
\end{equation}
which provides a tighter upper bound than those derived in the context of gravitational fields; see, for instance \cite{Scardigli_Casadio}. However, the upper bound \myref{Eq57} is weaker in comparison to \cite{Quesne_Tkachuk,Bawaj, Scardigli_Lambiase_Vagenas}, which were obtained in different contexts in the literature. Nevertheless, what we notice is that we obtain positive contributions from both SCQG and GUP deformation, which will make the theoretical values of $\delta$ closer to the experiment.
%%%%%%%%%%%%%%%%%%%%%%%%%%%%%%%%%%%%%%%%%%%%%%%%%%%%%%%%%%%%%%%%%%%%%%%%%%%%%%
%  Section 6
%%%%%%%%%%%%%%%%%%%%%%%%%%%%%%%%%%%%%%%%%%%%%%%%%%%%%%%%%%%%%%%%%%%%%%%%%%%%%%
\section{A schematic proposal for experiment}\label{sec6} 
To this end, we make a proposal for an experimental quantum bouncer through an opto-mechanical set-up, which would be able to provide an understanding of the combined effect of SCQG and GUP. Opto-mechanical devices yield a promising avenue for preparing and investigating quantum states of massive objects ranging from a few picograms up to several kilograms \cite{Kippenberg}. Significant experimental progress has been achieved by using such devices in different contexts, including coherent interactions \cite{Connell}, laser cooling of nano and micro-mechanical devices into their quantum ground state \cite{Teufel}. Recently, such types of systems have been utilized for the understanding of more exciting features like the SNE in SCQG \cite{Grossardt,Gan} and GUP \cite{Pikovski}. Here, our goal is to understand the combined effect of SCQG and GUP through the given opto-mechanical system.

The underlying principle behind the experiment is that the ultra cold neutrons (UCN) move freely in the gravitational field above a mirror and make a total reflection from the surface of the mirror when the corresponding de Broglie wavelength is bigger than the interatomic distances of the matter. Thus, the set-up gives rise to a GQW, where the UCN form bound quantum states in the Earth's gravitational field. The eigenvalues are non-equidistant and, are in the range of pico-eV. These type of scenarios offer fascinating possibilities to combine the effects of Newton's gravity at short distances with the high precision resonance spectroscopy methods of quantum mechanics.
\begin{figure}[ht]
\centering   \includegraphics[width=11.0cm,height=7.0cm]{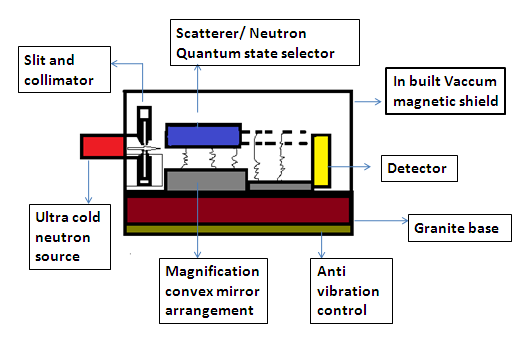}
\caption{The schematic of the opto-mechanical setup for the proposed experiment}
\label{Fig1}
\end{figure} 
The schematic diagram of the experimental set-up is given in Fig. \ref{Fig1}, which consists of an anti-vibrator granite table, a convex magnification system (mirror), an inclinometer, a ceiling scatterer and a position sensitive detector. The UCN shall be formed by passing the neutrons of the proposed wavelength between 0.75- 0.90 nm through the superfluid of \textsuperscript{4}He of a volume of about 4-5 liters. The UCN are brought to the experimental set-up by a guided system composed of a slit, a neutron collimator and an aperture. The neutrons fly in the slit and are screened in the range of 4-10 m/s. The upper wall of the slit is an efficient absorber which only lets the surviving neutrons to pass and, hence forth controlling first quantum states of the neutron flux. The mirror consists of a magnification system , which is made of a bi-metal (NiP) coated glass rod of radius between 3-3.5 nm with the roughness of about 1.5 nm. However the mirror system shall be modified to elliptical shape in order to harness the neutrons of much smaller wavelength
as well. The distribution of 100 $\mu$m in height is magnified to $\sim 2.5$ mm at the position on the detector. This gives the average magnification power of about 25 at the glancing angle at $20^0$ to make the reflection high.

The most important aspect of the position sensitive neutron detector is that  the neutrons must be changed into a charged particles by a nuclear reaction. The converter materials are available as \textsuperscript{3}He, \textsuperscript{6}Li, \textsuperscript{10}B, and \textsuperscript{157}Gd, etc. A black thin CCD (charge coupled detector) could be used along with a thin layer of \textsuperscript{3}He, \textsuperscript{10}B or  Ti-\textsuperscript{10}B-Ti. The ideal thickness of \textsuperscript{10}B should be about 200 nm, and must be evaporated directly on the  CCD surface. The standard pixel size is 24 $\mu$m $\times$ 24 $\mu$m and the thickness of the active volume should be about 20 $\mu$m. The  kinetic energy is deposited on the active area and a charge cluster is created, which typically spreads into nine pixels. The weighted center of the charge cluster is a good estimation of the incident neutron position.

At the exit of the UCN system the wavefucntion changes due to the absence of suppressing slit and as such the neutrons turn as bouncers on the mirror surface and their trajectories are measured by the Time of flight (TOF) method. They are characterized by a Gaussian distribution with a mean of 9.4 m/s and the standard deviation of 2.8 m/s which is subject to change due to various modifications of the design.
%%%%%%%%%%%%%%%%%%%%%%%%%%%%%%%%%%%%%%%%%%%%%%%%%%%%%%%%%%%%%%%%%%%%%%%%%%%%%%
%  Section 7
%%%%%%%%%%%%%%%%%%%%%%%%%%%%%%%%%%%%%%%%%%%%%%%%%%%%%%%%%%%%%%%%%%%%%%%%%%%%%%
\section{Conclusions}\label{sec7} 
We have studied a two-fold correction in the energy spectrum of a GQW. The first correction arises from the scheme of the SCQG itself, where we considered the SNE in a particular framework of braneworld model with infinite extra dimensions. The second correction emerges from a GUP deformation of the given SCQG framework. The combined effect of these two interesting theories provide a new bound on the energy spectrum of the GQW, which is closer to the observed values in the GRANIT experiment. Since, both of the effects of SCQG and GUP occur at small length scales, and both of the theories provide positive contributions in the correction of the energy spectrum, it raises a natural possibility that the combined theory may guide us towards a theory beyond the SCQG. Thus our proposal may yield an intermediate theory beyond the SCQG and the complete theory of quantum gravity. Moreover, we have provided a schematic experimental set-up which would help the laboratory experts to explore our theory further in the lab.

There are many natural directions that may follow up our investigation. First, it will be interesting to study the similar kind of effects of GUP in the context of other theories of gravity. There exist many other type of GUP deformation, which may be useful to study the similar theories to confirm our findings. However, the most interesting future problem lies on the understanding of the experimental realization of the combined theory of SCQG and GUP by using the opto-mechanical set-up, while it has already been used to understand each of the individual frameworks of SCQG and GUP. 

\vspace{0.5cm} \noindent \textbf{\large{Acknowledgements:}}  AB is supported by MHRD, Government of India and would like to thank Department of Physics and MMED at NIT Srinagar for carrying out her research pursuit. SD is supported by a CARMIN postdoctoral fellowship by IHES and IHP. The work of QZ is supported by NUS Tier 1 FRC Grant R-144-000-360-112.
%%%%%%%%%%%%%%%%%%%%%%%%%%%%%%%%%%%%%%%%%%%%%%%%%%%%%%%%%%%%%%%%%%%%%%%%%%%%%%
%  References
%%%%%%%%%%%%%%%%%%%%%%%%%%%%%%%%%%%%%%%%%%%%%%%%%%%%%%%%%%%%%%%%%%%%%%%%%%%%%%

%\bibliographystyle{unsrt} 
%\bibliography{Ref.bib}
%\input{Reference.tex}

%%%%%%%%%%%%%%%%%%%%%%%%%%%%%%%%%%%%%%%%%%%%%%%%%%%%%%%%%%%%%%%%%%%%%%%%%%%%%%
%  Rulling
%%%%%%%%%%%%%%%%%%%%%%%%%%%%%%%%%%%%%%%%%%%%%%%%%%%%%%%%%%%%%%%%%%%%%%%%%%%%%%
%\vspace{0.3cm}
%\begin{center}
%\rule{2.5cm}{1.0pt}
%\end{center}
%%%%%%%%%%%%%%%%%%%%%%%%%%%%%%%%%%%%%%%%%%%%%%%%%%%%%%%%%%%%%%%%%%%%%%%%%%%%%%

\end{document}